\documentclass[12pt,a4paper]{article} 
 \def\int2{\int\,\int}
 \def\int3{\int\,\int\,\int}
 \def\mint{\int\,\cdots\,\int}
 \usepackage{amsmath, amssymb}
 \usepackage{bm}
 \usepackage{graphicx}
\begin{document}
\textwidth=135mm
 \textheight=200mm
\begin{center}
{\bfseries Estimation of multi-particle correlation from multiplicity distribution 
observed in relativistic heavy ion collisions\footnote{{\small Talk presented at 
"VI Workshop on Particle Correlations and Fentscopy ", BITP, Ukraine, September 13 - 18, 2010.}}}
\vskip 5mm
 N. Suzuki$^a$, M. Biyajima$^b$ and T. Mizoguchi$^c$
\vskip 5mm
{\small {\it $^a$ Department of comprehensive management, Matsumoto University, Matsumoto 390-1295, Japan}} \\
{\small {\it $^b$ Department of physics, Shinshu University, Matsumoto 390, Japan}} \\
{\small {\it $^c$ Toba National College of Maritime Technology, Toba 517-8501, Japan  }}
\end{center}
\vskip 5mm
\centerline{\bf Abstract}

Analytical formula for multiplicity distribution is derived in the QO approach, 
where chaotic and coherent fields are contained. 
Observed charged multiplicity distributions in Au+Au collisions at $\sqrt{s}=200$ AGeV and in pp collisions  
at $\sqrt{s}=900$ GeV are analyzed by the formula. 
Chaoticity parameters in the inclusive events estimated from the analysis of multiplicity distributions are 
compared with those estimated from the analysis of observed two-particle inclusive identical particle correlations.

 \vskip 10mm
\section{\label{sec:intro}Introduction}

 In high energy nucleus-nucleus (AA) collisions or hadron-hadron collisions,  
 Bose-Einstein correlations of identical particles are considered as one of the possible measures 
 for the space-time domain where identical particles are produced.
As the colliding energy of AA collisions increases, thousands of identical particles can be produced in an event. 
Then, the production domain of those particles can be analyzed precisely event by event, 
or among the events with fixed multiplicity. 
Up to the present, most of theoretical approaches to identical particle correlations at fixed multiplicity are investigated 
  in the case of purely chaotic field~\cite{prat93, chao95, csor98, ledn00}. 

One of the theoretical approaches to the Bose-Einstein correlations is made on the analogy of the quantum optics
~\cite{glau63}, where two types of sources, chaotic and coherent are introduced.  
A diagrammatical method, based on the Glauber-Lachs formula~\cite{glau63}, has been proposed \cite{biya90} 
to find higher order Bose-Einstein correlation (BEC) functions.
In Ref.\cite{suzu97}, the generating functional (GF) for the momentum densities in the inclusive events is derived, 
and a diagrammatic representation for cumulants is proposed. 
Identical particle correlations at fixed multiplicity are formulated in \cite{suzu99}.

In the present paper, analytical formula for multiplicity distribution, which is identified to the Glauber-Lachs formula, is derived. 
A relation between the chaoticity parameter $p_{\rm sm}$ in the semi-inclusive events 
and that  $p_{\rm in}$ in the inclusive events is also obtained. 
Observed multiplicity distributions are analyzed by the Glauber-Lachs formula. 
Estimated value of $p_{\rm in}$ from the observed multiplicity distributions is compared with that from the observed inclusive identical two-particle correlations. 

\section{Generating functional}

In the QO approach, the $n$-particle momentum density 
in the semi-inclusive events is defined by, 
 \begin{eqnarray}
     \rho_n(p_1,\cdots,p_n) &=& c_0 \bigl\langle |f(p_1)|^2\cdots|f(p_n)|^2 \bigr\rangle_a,      \label{eq:gf1} \\ 
     f(p) &=& \sum_{i=1}^M a_i\phi_i(p) + f_c(p),   \label{eq:gf2}
 \end{eqnarray}
where $c_0$ is a normalization factor. In Eq.(\ref{eq:gf2}), 
$\phi_i(p)$ and $f_c(p)$ are amplitudes of the $i$-th chaotic source and a coherent source, respectively, 
and $a_i$ is a random complex number attached to the $i$-th chaotic source.  
The number $M$ of independent chaotic sources is regarded to be infinite. 

In Eq.(\ref{eq:gf1}), parenthesis $\langle F\rangle_a$ denotes an average of $F$ 
over the random number $a_i$ with a Gaussian weight~\cite{glau63}:
 \begin{eqnarray}
       \langle F \rangle_a
      = \prod_{i=1}^M \Bigl( {1 \over {\pi \lambda_i}} \int 
        \exp[-{\bigl| a_i \bigr|^2 \over \lambda_i}] d^2 a_i \Bigr) F.  \label{eq:gf5}
 \end{eqnarray}
The single-particle and the two-particle momentum densities are respectively given as
 \begin{eqnarray*}
      \rho_1(p_1)&=&c_0\langle |f(p_1)|^2\rangle_a  =c_0[r(p_1,p_1)+c(p_1,p_1)],  \\
      \rho_2(p_1,p_2)&=& c_0\langle |f(p_1)f(p_2)|^2\rangle_a  \nonumber \\
               &=&c_0\bigl\{ \rho(p_1)\rho(p_2)+|r(p_1,p_2)|^2
                +2{\rm Re}[r(p_1,p_2)c(p_2,p_1)] \bigr\},
 \end{eqnarray*}
where $r(p_1,p_2)$ is a correlation caused by the chaotic sources and 
$c(p_1,p_2)$ is a correlation by the coherent source,    
 \begin{eqnarray}
     r(p_1,p_2) = \sum_{i=1}^M \lambda_i \phi_i(p_1) \phi^*_i(p_2 ),     \quad
     c(p_1,p_2) = f_c(p_1) f^*_c(p_2).              \label{eq:gf6}   
 \end{eqnarray}

The GF of the semi-inclusive events is defined by the following equation,
 \begin{equation}
        Z_{\rm sm}[h(p)] = \sum_{n=1}^\infty {1 \over {n!}} 
         \mint \rho_n (p_1,\ldots,p_n) 
        h(p_1)\cdots h(p_n) {d^3p_1 \over E_1}\cdots {d^3p_n \over E_n}.       \label{eq:gf7}
 \end{equation}
From Eqs.(\ref{eq:gf1}) and (\ref{eq:gf7}), the GF is written as
 \begin{equation}
        Z_{\rm sm}[h(p)] = c_0 \Bigl\langle \exp \Bigl[
            \int\mid f(p)\mid^2 h(p) {d^3p \over E} \Bigr]  \Bigr\rangle_a,
                  \label{eq:gf8}
 \end{equation}
where an additional constant $Z_{\rm sm}[h(p)=0]$ is added to the right hand side of Eq.(\ref{eq:gf8}). 
Inversely, the $n$-particle momentum density in the semi-inclusive events is given from the GF as
 \begin{eqnarray*}
   \rho_n(p_1,\ldots,p_n)=\left. E_1\cdots E_n
         \frac{\delta^nZ_{\rm sm}[h(p)]}{\delta h(p_1)\cdots \delta h(p_n)}
         \right|_{h(p)=0}.
 \end{eqnarray*}
The generating function for multiplicity distribution $P(n)$ is given from Eq.(\ref{eq:gf7}) 
if function $h(p)$ is independent from momentum $p$:
 \begin{equation}
        Z_{\rm sm}(h) = c_0 \Bigl\langle \exp \Bigl[
            \int\mid f(p)\mid^2 {d^3p \over E} h \Bigr] 
              \Bigr\rangle_a.                   \label{eq:gf9}
 \end{equation}
The multiplicity distribution (MD) is given from Eq.(\ref{eq:gf9}) as
 \begin{eqnarray*}
    P(0)=  Z_{\rm sm}(0) = c_0, \quad P(n)=\frac{1}{n!}\left. \frac{\partial^n Z_{\rm sm}(h)}
            {\partial h^n} \right|_{h=0}.  
 \end{eqnarray*}
\section{Cumulants}

The GF $G_{\rm sm}[h(p)]$ for cumulants in the semi-inclusive events is defined by the following equation,
 \begin{equation}
     G_{\rm sm}[h(p)] \equiv \ln Z_{\rm sm}[h(p)]. \label{eq.cd3}
 \end{equation}
The $n$-th order cumulant is given by 
 \begin{equation}
      g_n(p_1,\ldots,p_n) = \left. E_1\cdots E_n
       \frac{\delta^n G_{\rm sm}[h(p)]}
        {\delta h(p_1)\cdots \delta h(p_n)}\right|_{h(p)=0}. 
                               \label{eq.cd4}
 \end{equation}
From Eqs.(\ref{eq:gf8}), (\ref{eq.cd3}) and (\ref{eq.cd4}), we have an iteration relation for momentum densities, 
 \begin{eqnarray}
    \rho_1(p_1)&=& c_0g_1(p_1),    \nonumber \\
    \rho_n(p_1,\ldots,p_n) 
       &=& g_1(p_1)\rho_{n-1}(p_2,\ldots,p_n)  \nonumber \\
       &+& \sum_{i=1}^{n-2} \sum 
               g_{i+1}(p_1,p_{j_1},\ldots,p_{j_i}) 
        \rho_{n-i-1}(p_{j_{i+1}},\ldots,p_{j_{n-1}}) \nonumber \\
       &+& c_0g_n(p_1,\ldots,p_n).     \label{eq.cd5}
 \end{eqnarray}
The second summation on the right hand side of Eq.(\ref{eq.cd5}) indicates 
that all possible combinations of $(j_1,\cdots, j_i)$ 
and $(j_{i+1},\ldots,j_{n-1})$ are taken from $(2,3,\ldots, n)$. 
The $n$-particle momentum density $\rho_n(p_1, \ldots,p_n)$ $(n=1,2,\ldots)$ is written by the cumulant  
$g_i(p_1,\ldots ,p_i)$ $(i=1,2,\ldots ,n)$ from Eq.(\ref{eq.cd5}).

Diagrammatic representation of Eq.(\ref{eq.cd5}) is shown in \cite{suzu99}.  
For example, cumulants up to the third order are written explicitly as
 \begin{eqnarray}
      g_{12}  &=& r_{12}r_{21} + c_{12}r_{21}+r_{12}c_{21}, \nonumber \\
      g_{123} &=& r_{12}r_{23}r_{31}
                 + c_{12}r_{23}r_{31}
                 + r_{12}c_{23}r_{31}
                 + r_{12}r_{23}c_{31} + c.c.,                       \label{eq.cd6}
 \end{eqnarray}
where the following abbreviations are used,
 \begin{eqnarray*}
     g_{j_1\cdots j_m} = g_m(p_{j_1},\ldots ,p_{j_m}),   \quad
     r_{ij} &=& r(p_i,p_j),\quad c_{ij} = c_(p_i,p_j), 
 \end{eqnarray*}
and complex conjugates to the terms explicitly shown are denoted by c.c..

\section{Multiplicity distribution}

From Eqs.(\ref{eq.cd5}) and (\ref{eq.cd6}), we obtain a recurrence equation for MD,  
 \begin{eqnarray}
     P(n) 
     =  \frac{1}{n} \sum_{j=1}^n \bigl[ \Delta_j^{(R)} + j \Delta_{j-1}^{(S)} \bigr]P(n-j), \quad n=1, 2, \ldots,      \label{eq.cd9}
 \end{eqnarray}
where, with $R_0(k_1,k_2) = \omega_1 \delta^3(\bm{k}_1-\bm{k}_2)$, 
 \begin{eqnarray}
 && \Delta_j^{(R)} = \int R_j(k,k)\frac{d^3k}{\omega},  \,\,\,
    R_j(p_1,p_2)= \int r(p_1,k)R_{j-1}(k,p_2)\frac{d^3k}{\omega},   \label{eq.cd13} \\
 && \Delta_{j-1}^{(S)} = \int S_{j-1}(k,k)\frac{d^3k}{\omega}, \,\,\,  
   S_{j-1}(p_1,p_2) = \int c(p_1,k)R_{j-1}(k,p_2)\frac{d^3k}{\omega}.  \nonumber \\   \label{eq.cd14}
 \end{eqnarray}

In the followings, variables are changed from $(p_{iL},\bm{p}_{iT})$ to $(y_i,\bm{p}_{iT})$, 
where $y_i$ is the rapidity of particle $i$.
Correlations $r(p_1,p_2)$ and $c(p_1,p_2)$ are both assumed to be real, and parametrized as~\cite{suzu99},
 \begin{eqnarray*}
     r(y_1,\bm{p}_{1T};y_2,\bm{p}_{2T}) &=& p_{\rm sm}
      \sqrt{\rho(y_1,\bm{p}_{1T})\rho(y_2,\bm{p}_{2T})}
                    \,I(\Delta y, \Delta \bm{p}_{1T}), 
                          \nonumber  \\
     c(y_1,\bm{p}_{1T};y_2,\bm{p}_{2T}) &=& (1-p_{\rm sm})
          \sqrt{\rho(y_1,\bm{p}_{1T})\rho(y_2,\bm{p}_{2T})}, \nonumber  \\            
      \rho(y_1,\bm{p}_{1T}) &=&  \langle n_0 \rangle \frac{\alpha^{1/2}\beta}{\pi^{3/2}}   
       \exp[ -\alpha\,y_1^2 - \beta\, \bm{p}_{1T}^2 ],  \nonumber  \\
    I(\Delta y, \Delta\bm{p}_T) &=& \exp[ -\gamma_L (\Delta y)^2 - \gamma_T(\Delta \bm{p}_T)^2 ],       \label{bec48}
 \end{eqnarray*}
where $\Delta y =y_2 -y_1$, $\Delta \bm{p}_T = \bm{p}_{2T} - \bm{p}_{1T}$, and $p_{\rm sm}$ is the chaoticity parameter 
in the semi-inclusive events. It is assumed to be constant at present. 

Then, $R_j(p_1,p_2)$ in Eqs.(\ref{eq.cd13}) and (\ref{eq.cd14}) is given by the following form,
 \begin{eqnarray*}
  R_j(y_1,\bm{p}_{1T},y_2,\bm{p}_{2T})&=& N_j\exp [-A_j(y_1^2+y_2^2)+2C_jy_1y_2]  \\
       &\times&  \exp [ - U_j(\bm{p}_{1T}^2+\bm{p}_{2T}^2) + 2W_j\bm{p}_{1T}\bm{p}_{2T}],  
 \end{eqnarray*}
and
 \begin{eqnarray}
  && \Delta_1^{(R)} = p_{\rm sm}\langle n_0 \rangle, \quad       
     \Delta_j^{(R)} = N_j \Bigl(\frac{\pi}{2(A_j-C_j)}\Bigr)^{1/2} \, \frac{\pi}{2(U_j-W_j)},  \label{eq.cd21} \\
  &&\Delta_0^{(S)} = (1-p_{\rm sm})\langle n_0 \rangle, \quad 
    \Delta_j^{(S)} = N_j \frac{\pi^{3/2}(1-p_{\rm sm})\langle n_0  \rangle }{\sqrt{A_j+A_1}\,(U_j+U_1)},  \label{eq.cd22}
 \end{eqnarray}
where
 \begin{eqnarray}
     &&  A_1 = \frac{\alpha}{2} + \gamma_L, \quad  A_{j+1} = A_1 - {\gamma_L^2}/({A_j+\alpha/2+\gamma_L}),  \label{eq.cd23} \\
     &&  C_1 = \gamma_L,  \quad   C_{j+1} = {\gamma_LC_j}/({A_j+\alpha/2+\gamma_L}),  \nonumber \\
     &&  U_1 = \frac{\beta}{2}+\gamma_T, \quad  U_{j+1} = U_1 - {\gamma_T^2}/({U_j+\beta/2+\gamma_T}), \label{eq.cd25} \\
     &&  W_1 = \gamma_T, \quad
       W_{j+1} = {\gamma_TW_j}/({U_j+\beta/2+\gamma_T}),  \nonumber  \\
      &&  N_1 = p_{\rm sm} \langle n_0 \rangle \frac{\alpha^{1/2}\beta}{\pi^{3/2}}, \quad
      N_{j+1} = \frac{p_{\rm sm} \langle n_0 \rangle \alpha^{1/2}\beta}
        {\sqrt{A_j+\alpha/2+\gamma_L}(U_j+\beta/2+\gamma_T)}N_j.   \nonumber
 \end{eqnarray}
Recurrence equations, (\ref{eq.cd23}) and (\ref{eq.cd25}) can be solved~\cite{csor98}.  
Let $F_j$ be defined by 
 \begin{eqnarray}
    F_{j}=A_{j+1} + A_1, \quad  j=0,1,2, \ldots,  \label{eq.cfra5}
 \end{eqnarray}
then, $F_j$ is given by a finite continued fraction~\cite{khov63}:
 \begin{eqnarray}
     F_j &=& {P_j}/{Q_j}, \quad j=0, 1, 2, \ldots,  \label{eq.cfra6}   \\
     P_j &=& q_0 P_{j-1} + q_1 P_{j-2}, \quad j=2,3, \ldots,  \label{eq.cfra7}   \\
     Q_j &=& q_0 Q_{j-1} + q_1 Q_{j-2}, \quad j=2,3,\cdots.  \label{eq.cfra8}  
 \end{eqnarray}
where, $q_0=\alpha + 2\gamma_L$, $q_1=-{\gamma_L}^2$, 
$P_0=q_0$, $Q_0 = 1$, $P_1={q_0}^2 + q_1$Cand $Q_1= q_0$. 
Therefore, $P_j$ and $Q_j$ are given respectively by
 \begin{eqnarray*}
       P_j = ({{r_2}^{j+2}  - {r_1}^{j+2} })/({r_2 - r_1}),  \quad                              
       Q_j = ({{r_2}^{j+1}  - {r_1}^{j+1} })/({r_2 - r_1}),                              
 \end{eqnarray*}
where
 \begin{eqnarray*}
        r_1 = \frac{\alpha +2\gamma_L - \sqrt{\alpha^2 + 4 \alpha\gamma_L}}{2}, \quad  
        r_2 = \frac{\alpha +2\gamma_L + \sqrt{\alpha^2 + 4 \alpha\gamma_L}}{2}.                             
 \end{eqnarray*}
Then, $A_j$ and $C_j$ in Eqs.(\ref{eq.cd21}) and (\ref{eq.cd22}) are written respectively as,
 \begin{eqnarray}
     A_j &=& \frac{r_2-r_1}{2} \frac{ 1 + (r_1/r_2)^j}{ 1 - (r_1/r_2)^j }, \quad
     C_j = (r_2-r_1)  \frac{(r_1/r_2)^{j/2}}{ 1 - (r_1/r_2)^j }.  \label{eq.cfra10}
 \end{eqnarray}
%
Similarly, $U_j$ and $W_j$ are respectively given by
 \begin{eqnarray} 
     U_j &=& \frac{1}{2}(t_2-t_1) \frac{ 1 + (t_1/t_2)^j}{ 1 - (t_1/t_2)^j },  \quad
     W_j = (t_2-t_1)  \frac{(t_1/t_2)^{j/2}}{ 1 - (t_1/t_2)^j },     \label{eq.cfra11}  
 \end{eqnarray}
where 
 \begin{eqnarray*}      
     t_1 = \frac{ \beta + 2\gamma_T - \sqrt{\beta^2 + 4 \beta\gamma_T} }{2}, \quad
     t_2 = \frac{ \beta + 2\gamma_T + \sqrt{\beta^2 + 4 \beta\gamma_T} }{2}.                                     
 \end{eqnarray*}
From Eqs.(\ref{eq.cd9}) and (\ref{eq.cfra11}), following expressions are obtained:
 \begin{eqnarray} 
    \Delta_{j}^{(R)} &=& \xi^{j} \{ 1-(r_1/r_2)^{j/2} \}^{-1}
              \bigl\{ 1-(t_1/t_2)^{j/2} \bigr\}^{-2},   \label{eq.cfra12}\\
    \Delta_{j-1}^{(S)} &=& A_0 \xi^{j-1} \bigl\{ 1-(r_1/r_2)^j \bigr\}^{-1/2} 
              \bigl\{ 1-(t_1/t_2)^j \bigr\}^{-1},   \label{eq.cfra13}
 \end{eqnarray}
where
 \begin{eqnarray}
        \xi &=& \frac{p_{\rm sm} \langle n_0 \rangle \sqrt{\alpha}\beta }{\sqrt{r_2}\,t_2}
              = \bigl( 1 - \sqrt{r_1/r_2}\, \bigr) 
                \bigl( 1 - \sqrt{t_1/t_2}\, \bigr)^2 p_{\rm sm} \langle n_0 \rangle, \label{eq.cfra14} \\    
        A_0 &=& \sqrt{ 1-{r_1}/{r_2} }\,\bigl( 1-{t_1}/{t_2} \bigr)(1-p_{\rm sm})\langle n_0 \rangle. \label{eq.cfra15}
 \end{eqnarray}

The generating function for multiplicity distribution $P(n)$ is given by
 \begin{eqnarray*}
   \Pi(z) = Z_{\rm sm}(z) = \sum_{j=0}^{\infty} P(n) z^n.
 \end{eqnarray*}
Then the differential equation for $\Pi(z)$ is obtained from Eq.(\ref{eq.cd9}) as,
 \begin{eqnarray}
        \frac{d}{dz}\Pi(z) = 
           \sum_{j=1}^{\infty} \Bigl( \Delta_j^{(R)} + j \Delta_{j-1}^{(S)} \Bigr)z^{j-1} \Pi(z).   \label{eq.cfra16} 
 \end{eqnarray}
For $j>>1$, Eqs.(\ref{eq.cfra12}) and (\ref{eq.cfra13}) can be approximated as $ \Delta_j^{(R)} \simeq  \xi^j$ and 
$\Delta_{j-1}^{(S)} \simeq A_0 \xi^{j-1}$.
Therefore, with the boundary condition $\Pi(1)=1$, we obtain 
 \begin{eqnarray}
    \Pi(z) =  \bigl\{ 1- p_{\rm in} \langle n \rangle(z-1) \bigr\}^{-1}
         \exp \Bigl[(1 - p_{\rm in}) \langle n \rangle\frac{(z-1)}{1- p_{\rm in} \langle n \rangle(z-1)} \Bigr],  \label{eq.cfra19}  
 \end{eqnarray}
where the average multiplicity $\langle n \rangle$ and the chaoticity parameter $p_{\rm in}$  in the inclusive events are given by the following equations,
 \begin{eqnarray}
    \langle n \rangle = {\xi}/({1-\xi}) + {A_0}/{(1-\xi)^2}, \quad 
     p_{\rm in} \langle n \rangle = {\xi}/({1-\xi}).      \label{eq.cfra18}      
 \end{eqnarray}

The multiplicity distribution $P(n)$ is given from Eq.(\ref{eq.cfra19}) as 
 \begin{eqnarray}
    P(n) &=& \frac{1}{n!} \frac{\partial^n}{\partial z^n}\Pi(z)\Big|_{z=0}  \\
         &=&   \frac{ (p_{\rm in} \langle n \rangle)^n }{ (1 + p_{\rm in} \langle n \rangle)^{n+1} } 
         \exp \Bigl[-\frac{(1 - p_{\rm in}) \langle n \rangle}{1+ p_{\rm in} \langle n \rangle}\Bigr]
         L_n\Bigl( \frac{(1- p_{\rm in}) \langle n \rangle }{1+p_{\rm in} \langle n \rangle } \Bigr),  \label{eq.cfra20}  
 \end{eqnarray}
where $L_n(x)$ denotes the Laguerre polynomial. 
Equation (\ref{eq.cfra20}) is called the Glauber-Lachs formula~\cite{glau63, nami75}.
The KNO scaling function of  the Glauber-Lachs formula is given by,
 \begin{eqnarray}
   \phi(z) = \frac{1}{p_{\rm in}} 
            \exp \Bigl[-\frac{ z + 1 - p_{\rm in} }{ p_{\rm in} } \Bigr]
         I_0\Bigl( \frac{2}{ p_{\rm in} } \sqrt{(1-p_{\rm in}) z }\, \Bigr),     \label{eq.cfra21}  
 \end{eqnarray}
where $I_0(z)$ is the modified Bessel function.

\section{Analysis of experimental data}

\begin{figure}[!htb]
  \begin{center}
  \begin{minipage}{0.45\linewidth}
   \begin{center}
   \includegraphics[width=\linewidth]{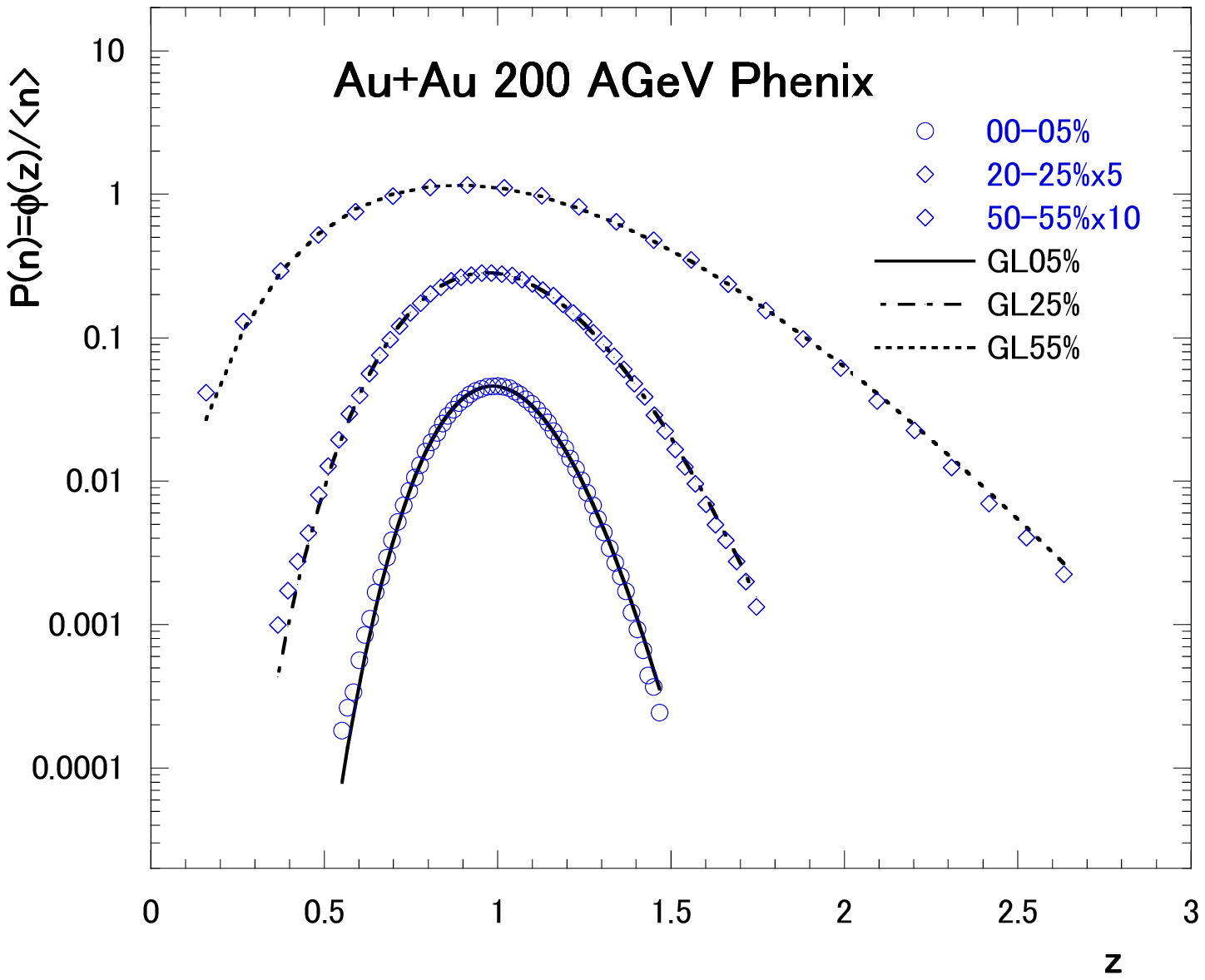}
   \caption{Analysis of charged multiplicity distributions observed in Au+Au collisions~\cite{adar08} by Eq.(\ref{eq.cfra22})}\label{fig.au200phen}
   \end{center}
  \end{minipage}  
  \hspace{4mm}
  \begin{minipage}{0.45\linewidth}
   \begin{center}
   \includegraphics[width=\linewidth]{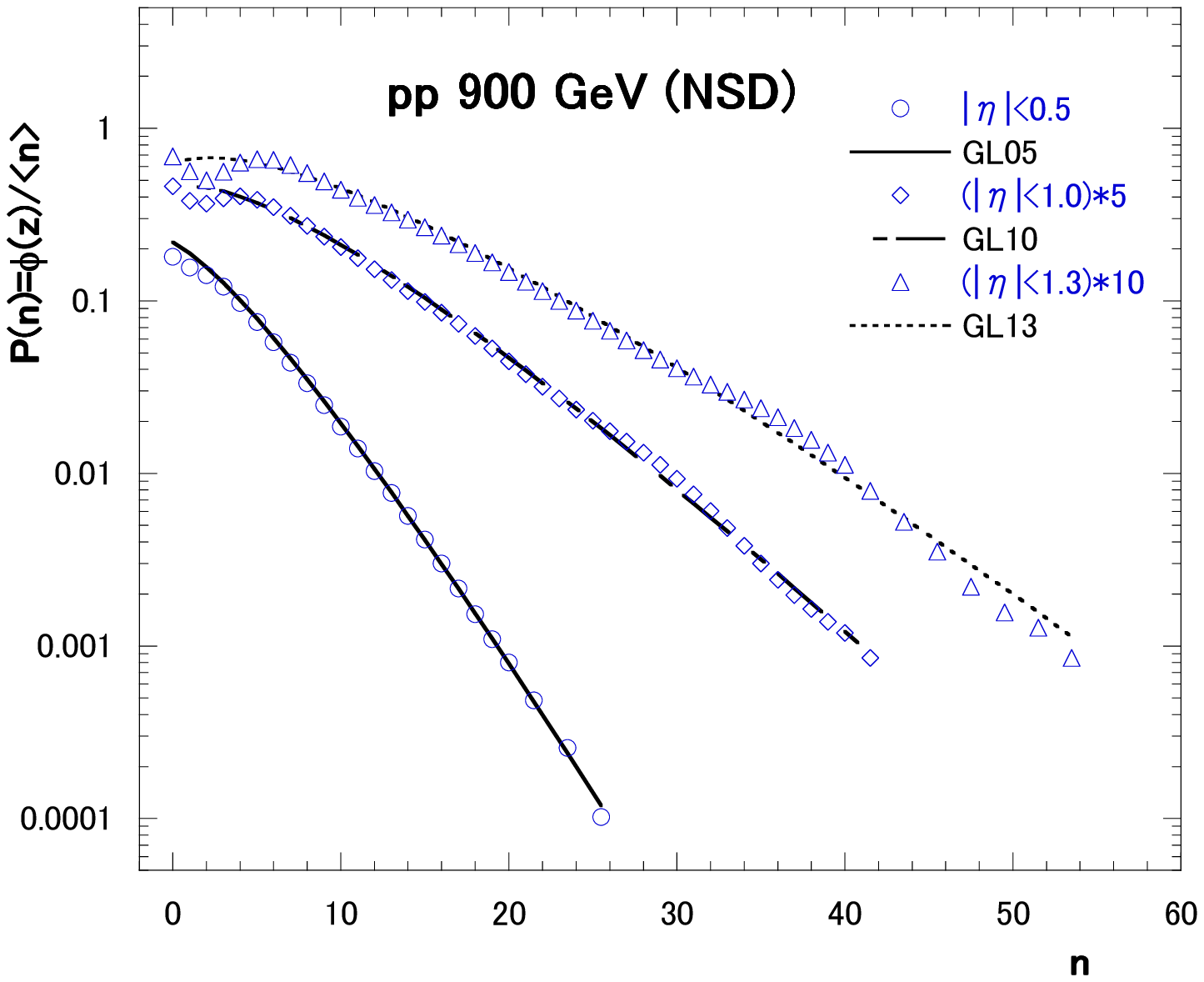}
   \caption{Analysis of charged multiplicity distributions observed in pp collisions~\cite{aamo10a}  by Eq.(\ref{eq.cfra22})}\label{fig.pp900alice}
   \end{center}
  \end{minipage} 
  \end{center} 
 \end{figure}

At first, observed multiplicity distributions in Au+Au collisions at $\sqrt{s} = 200$ AGeV~\cite{adar08} are analyzed by the scaling form
 \begin{eqnarray}
   P(n) = \phi(z)/\langle n \rangle, \quad z = n / \langle n \rangle.     \label{eq.cfra22}  
 \end{eqnarray}
Results for Au+Au collisions and the estimated parameters from the analysis are shown in Fig.\ref{fig.au200phen} and Table \ref{tab.au200phenix}, respectively.  

The second order BEC function in the QO approach for the Gaussian source function is given by
 \begin{eqnarray*}
    N^{2-}/N^{BG} = 1 + 2p_{\rm in}(1-p_{\rm in})E_{2B} + {p_{\rm in}}^2{E_{2B}}^2, \quad
    E_{2B} = \exp[-r^2{q_{inv}}^2]. 
 \end{eqnarray*}
At $q_{inv}=0$,  $N^{2-}/N^{BG} - 1= p_{\rm in}(2-p_{\rm in})$. It corresponds to correlation strength $\lambda$.

The correlation strength $\lambda$, estimated from the experimental data with full Coulomb corrections 
for $0.2\,\, {\rm GeV/c} < p_T < 2.0\,\, {\rm GeV/c}$ 
for 0-30\% centrality, is about 0.39~\cite{adle04}, which is much larger than $p_{\rm in}(2-p_{\rm in})$
in Table \ref{tab.au200phenix} estimated from the multiplicity distributions.

 \begin{table}[!htb] 
    \caption{Estimated parameters in the analysis of charged multiplicity distributions observed 
    in Au+Au collisions at $\sqrt{s}=200$ AGeV~\cite{aamo10a}.}  
   \label{tab.au200phenix}    
 \begin{center}
  \begin{tabular}{ccccc}
     Centrality  & $\langle n \rangle \pm \delta \langle n \rangle$ & $p_{\rm in} \pm \delta p_{\rm in}$ & $\chi_{\rm min}^2/n.d.f$ 
     &  $p_{\rm in}(2-p_{\rm in})$ \\ \hline
     00 - 05\%  &   62.3 $\pm$ 0.04   &  $( 9.81 \pm 0.07 )\times 10^{-3}$  & 879.3/(58-2) & 0.0195 \\
     20 - 25\%  &   34.0 $\pm$ 0.03  &  $( 2.19 \pm 0.01 )\times 10^{-2}$  & 475.9/(48-2) & 0.0433 \\ 
     50 - 55\%  &    9.31 $\pm$ 0.03  &  $( 7.52 \pm 0.09 )\times 10^{-2}$  & 914.7/(24-2) & 0.145 \\ \hline
  \end{tabular}  
 \end{center} 
 \end{table} 

Then observed multiplicity distributions in pp collisions at $\sqrt{s}=900$ GeV~\cite{aamo10a} are analyzed by Eq.(\ref{eq.cfra22}).
Results and estimated parameters are shown in Fig.\ref{fig.pp900alice} and Table \ref{tab.pp900alice}, respectively. 

The correlation strength $\lambda$, estimated from the experimental data for $0.1\,\, {\rm GeV/c} < p_T < 0.25\,\, {\rm GeV/c} $
for $|\eta|<0.8 $, is $0.628 \pm 0.133$~\cite{aamo10b}. It is not inconsistent with $p_{\rm in}(2-p_{\rm in})=0.729$ 
in Table \ref{tab.pp900alice} estimated from the multiplicity distributions for pseudorapidity range $|\eta|<1.0$.

 \begin{table}[!htb] 
   \caption{Estimated parameters in the analysis of charged multiplicity distributions observed in pp collisions at $\sqrt{s}=900$ GeV~\cite{aamo10a}.}  
   \label{tab.pp900alice}  
 \begin{center}
  \begin{tabular}{ccccc}
     $\eta$ range & $\langle n \rangle \pm \delta \langle n \rangle$ & $p_{\rm in} \pm \delta p_{\rm in}$ & $\chi_{\rm min}^2/n.d.f$ 
     & $p_{\rm in}(2-p_{\rm in})$  \\ \hline
     $ |\eta|<0.5 $ &   3.82 $\pm$ 0.05  &  $ 0.579 \pm 0.03$    & 45.2/(24-2)  &  0.823 \\   
     $ |\eta|<1.0 $ &   7.59 $\pm$ 0.06  &  $ 0.479 \pm 0.01$    & 48.5/(42-2)  &  0.729 \\ 
     $ |\eta|<1.3 $ &   9.92 $\pm$ 0.08  &  $ 0.434 \pm 0.01$    & 71.8/(48-2)  &  0.680 \\ \hline
  \end{tabular}  
 \end{center}    
 \end{table} 

\section{Summary and discussions}

The analytical formula for multiplicity distribution in the QO approach is derived. It becomes the Glauber-Lachs formula.  
A relation between the chaoticity parameter $p_{\rm in}$ in the inclusive events and that $p_{\rm sm}$  in the semi-inclusive events is obtained.
Multiplicity distributions observed in the Au+Au collisions at $\sqrt{s}=200$ AGeV and in pp collisions at $\sqrt{s}=900$ GeV are analyzed by the scaling form of Glauber-Lachs formula. The correlation strength calculated with $p_{\rm in}$ is compared with that measured from the second order BEC data. 
   
In Au+Au collisions, the former calculated with $p_{\rm in}$ is much smaller than the latter measured from the BEC data. It would be caused by the fact that the centrality region for the data sample of MD is different from that for the second order BEC. It would be very interesting, whether both values are consistent or not, if the MD and the second order BEC are taken from the same centrality region.

In pp collisions, the former is not inconsistent with the latter.

   %

%

\begin{thebibliography}{99}
%
 \bibitem{prat93}\textit{Pratt S.} // Phys. Lett. B.  1993. V.301. P.159.
%
 \bibitem{chao95}\textit{Chao W. Q., Gao C. S., Zhang Q. H.} // J. Phys. G. 1995. V.21. P.847;
                 \textit{Zhang Q. H.} // Phys. Lett. B. 1997. V.406. P.159.
%
 \bibitem{csor98} \textit{Cs\"{o}rg\H{o} T., Zimanyi J.} // Phys. Rev. Lett. 1998. V.80. P.916;  
     \textit{Zimanyi J., Cs\"{o}rg\H{o} T.} // Heavy Ion Phys. 1999. V.9. P.241.
%
 \bibitem{ledn00} \textit{Lednicky R. et al.} // Phys. Rev. C. 2000. V.61. 034901.
%
 \bibitem{glau63}\textit{Glauber R. J.}// Phys. Rev. 1963. V.131. P.2766;
     \textit{Lachs G. }// Phys. Rev. B. 1965. V.138. P.1012. 
%
 \bibitem{biya90}\textit{Biyajima M., et al.}// Prog. Theor. Phys. 1990. V.84. P.931;
     ibid. 1992. V.88. P.157.
%
 \bibitem{suzu97}
   \textit{Suzuki N., Biyajima M., Andreev I. V.}// Phys. Rev. C. 1997. V.56. P.2736;
   \textit{Suzuki N., Biyajima N.}// Prog. Theor. Phy. 1992. V.88. P.609. 
%
 \bibitem{suzu99}\textit{Suzuki N., Biyajima M.}// Phys. Rev. C. 1999. V.60. P.034903;
  Proc. of the 8-th International Symposium on Multiparticle Production 
     "Correlations and Fluctuations '98", Matrahaza, Hungary, June 14-21 1998. P.98.
     Eds. T.Cs\"{o}rg\H{o}, S.Hegyi, R.Hwa and G.Jancs\'{o}, World Scientific.
%
 \bibitem{khov63} \textit{Khovanskii A. N.} The application of continued fractions 
 and their generalizations to problems in approximation theory. Groningen, The Netherlands:  P. Noordhoff N. V., 1963.
%
 \bibitem{nami75}\textit{Namiki M., Ohba I., Suzuki N.} // Prog. Theor. Phys. 1975. V.53. P.775; 
 \textit{Biyajima M., Suzuki N.} // Prog. Theor. Phys. 1985. V.73. P.918.
%
 \bibitem{biya78}\textit{Biyajima M., Miyamura O., Nakai T.} // 
  Proc. of the Multiparticle Dynamics, Hakone, Japan, 1978. P.139
%
 \bibitem{adar08}\textit{Agare A. et al. (PHENIX Collab.)} // Phys. Rev. C. 2008. V.78. P.044902.
%
 \bibitem{adle04}\textit{Adler S.S. et al. (PHENIX Collab.)} // Phys. Rev. Lett. 2004. V.93. P.152302.
%
 \bibitem{aamo10a}\textit{Aamodt K. et al. (ALICE Collab.)} //  arXiv1004.3034[hep-ex].
%
 \bibitem{aamo10b}\textit{Aamodt K. et al. (ALICE Collab.)} //  arXiv1007.0516[hep-ex].
%
\end{thebibliography}
\end{document}